\documentclass[9pt,twocolumn,twoside]{osajnl}
\def\lsim{\raise0.3ex\hbox{$<$\kern-0.75em\raise-1.1ex\hbox{$\sim$}}}
\def\gsim{\raise0.3ex\hbox{$>$\kern-0.75em\raise-1.1ex\hbox{$\sim$}}}
\journal{ao} 

\setboolean{shortarticle}{false} 

\title{Common-mode rejection \\ in Martin-Puplett spectrometers \\ for astronomical observations at mm-wavelengths}

\author[1,*]{Giuseppe D'Alessandro}
\author[1]{Paolo de Bernardis}
\author[1]{Silvia Masi}
\author[1,**]{Alessandro Schillaci}

\affil[1]{Department of Physics, Sapienza University of Rome, Rome, IT}

\affil[*]{Corresponding author: giuseppe.dalessandro@roma1.infn.it}
\affil[**]{current address: Department of Physics, Jadwin Hall, Princeton University, Princeton, NJ 08544-0708 }
\dates{Compiled \today}

\ociscodes{(120.6200) Spectrometers and spectroscopic instrumentation ; (300.6300) Spectroscopy, Fourier transforms; (300.6270) Spectroscopy, far infrared; (350.1270) Astronomy and astrophysics.}

\doi{\url{http://dx.doi.org/10.1364/ao.XX.XXXXXX}}

\begin{abstract}
The Martin-Puplett interferometer (MPI) is a differential Fourier transform spectrometer (FTS), measuring the difference between spectral brightness at two input ports. This unique feature makes the MPI an optimal zero instrument, able to detect small brightness gradients embeddend in a large common background. In this paper we investigate experimentally the common-mode rejection achievable in the MPI at mm wavelengths, and discuss the use of the instrument to measure the spectrum of cosmic microwave background (CMB) anisotropy. \\ \\

\emph{One print or electronic copy may be made for personal use only. Systematic reproduction and distribution, duplication of any material in this paper for a fee or for commercial purposes, or modifications of the content of this paper are prohibited.}
\end{abstract}
\setboolean{displaycopyright}{true}
\begin{document}

\maketitle
\thispagestyle{fancy}
\ifthenelse{\boolean{shortarticle}}{\abscontent}{}

\section{Introduction}

Martin-Puplett interferometers \cite{Mart70} have been used widely in the laboratory and in astrophysical observations because, at variance with dispersion spectrometers, thery are intrinsically imaging, and, at variance with Fabry-Perot spectrometers, they are intrinsically differential. Moreover, the frequency coverage of a MPI is much wider than the free spectral interval of dipersion and Fabry Perot instruments. The MPI can be used in a differential configuration where a sky field is compared to an internal reference blackbody (as in the COBE-FIRAS instrument \cite{Math93}) or to a combination of blackbodies (as in the Herschel-SPIRE \cite{Grif07}). Both instruments have been very successful, due to the ability to perform accurate null-measurements (see e.g. \cite{Math90, Grif10}).  In the case of FIRAS, it has been possible to compare the brightness of the Cosmic Microwave Background (CMB) to the brightness of an internal reference blackbody at the same temperature (2.725K), placing an upper limit to the difference of the two brightnesses as low as 0.01\% of the peak brightness. This means that the rejection of the common mode signal is better than 1 part in 10000.

This feature of the MPI opens the possibility of its use as a polarimeter able to measure a tiny CMB polarization degree, as in the proposed PIXIE experiment (see \cite{Kogu11}).  The experiment can also be used as an accurate absolute spectrometer, closing one of the input ports with a calibrator, as was done with the FIRAS experiment, but improving in accuracy and sensitivity, so that tiny deviations expected in the CMB spectrum, due to energy injections in the early universe or to later intercations (see e.g. \cite{SZ69,Chlu12}) can be detected.

In a different configuration of the MPI, two sky-fields (or two orthogonal polarizations) are compared. This is obtained by placing the two input ports of the MPI in different locations of the focal plane of the telescope. The success of this configuration is closely linked to the extent that the instrument is able to reject common-mode signals coming from the instrument itself, the atmosphere, the foregrounds, and the CMB (see e.g. \cite{debe12, debe12a, Kogu11}).  The ratio between the output signal resulting from the difference in the two input signals and the small residual output signal resulting from the signal common at the two inputs is called the Common Mode Rejection Ratio (CMRR), and is usually given in dB. The larger the CMRR, the better the differential instrument is performing. 

The typical application for this configuration is the measurement of the Sunyaev-Zeldovich effect (see e.g. \cite{debe12a}), where the instrument has to compare the brightness coming from two lines of sight, one through the target cluster of galaxies, and one from a reference region outside the cluster. The brightness difference is of the order of $< 0.01\%$ of the common CMB brightness, and of $< 0.1\%$ of the emission of a warm metallic mirror.  Extremely high CMRR levels are thus required to obtain clean measurements of the SZ spectrum. On the other hand, using any other spectrometer type the required dynamic range would be extremely wide, so the MPI solution appears very appealing.

Several other scientific objectives can be explored using this configuration. A partial list includes the study of interstellar cooling lines in our Galaxy and in other galaxies in the sub-mm/mm range (from C$^+$, N$^+$, CO, etc., see e.g. \cite{Benn94} ); the study of the anisotropy of the Cosmic Infrared Background to sample the evolution of large scale structures in the Universe(see e.g. \cite{Vier13}); the effects of Rayleigh scattering on CMB photons through the spectrum of CMB anisotropy (see e.g. \cite{Yu01, Lewi13}); the study of  patchy reionization and in general of the WIHM through measurements of the patchy kinematic SZ effect (see e.g. \cite{Shao11}).

A few other spectrometer configurations have been conceived for similar scientific targets. Good examples are the z-spec instrument, a waveguide-coupled diffraction grating spectrometer described in \cite{Hina08}, and the filter-bank waveguide spectrometer described in \cite{Brya15}.  Both instruments rely on sky scans and AC-coupled detectors to remove the common mode. In a differential MPI, instead, the common-mode signal is not modulated by the moving mirror scan, and can easily be removed from the measured interferograms.

Given these very interesting but also very demanding applications, it is important to review the limiting factors for the CMRR in the MPI. 

The input section of the MPI is the most important in this respect. In fact, after the input wire-grid the two beams from the two inputs follow a common path in the instrument and are processed exactly in the same way. The input section includes two collimators (one for each input port) and one input wire-grid polarizer, transmitting radiation from one input and reflecting radiation form the other input. Unbalance in the two input throughputs, or a difference between the transmissivity and the reflectivity of the wire-grid, result inevitably in a non-zero signal even when the two sources at the input have the same brightness. This is an evident source of common mode signal, which can be minimized with a careful design of the input section. 

The second source of mismatch is more subtle and does not depend on the unbalance of the input section, but rather on the unperfect cancelling of perfectly equalized input signals inside the MPI. In the following we will call this second component as the {\sl real} common mode signal. The beamsplitter polarizer is important in this respect. In fact, the two input signals, after the input polarizer, have orthogonal polarizations. If the subsequent beamsplitter wire-grid is not properly oriented (i.e. if its main axis is not seen at 45$^o$ from the two incoming polarizations), another offset signal is produced, in addition to the previously described one, when the two input sources have the same brightness. We are studing this effect with dedicated measurements \cite{DAless2015}. 

Other unbalance factors are difficult to model, so it is interesting to investigate experimentally the CMRR achievable in a real instrument, and its dependence on the actual implementation.

To this purpose we have setup a custom experiment, using the OLIMPO DFTS (Differential Fourier Transform Spectrometer) MPI ( \cite{2014A&A...565A.125S} ) and filling the two input ports with radiation from two blackbody sources. For this test we have used room-temperature blackbodies: in this way the Signal to Noise Ratio (SNR) of the residual signal , coming from imperfect cancellation of the two input brightnesses, can be measured accurately, even for high values of the CMRR. 

In this paper we describe the setup and the results of the measurements, setting a lower limit of $\sim 50dB$ for the CMRR of our MPI.  We also discuss the suitability of this instrument for spectral measurements of the Sunyaev-Zeldovich (SZ) effect.

\section{Common mode rejection and SZ effect observations}

An important feature of all the instruments measuring differences between two input signals is the ability to eliminate the part of the signal which is common to the two inputs, and detect only the difference, even if the latter is much smaller than the former.  This is very difficult when the difference is much smaller than the two original inputs: a large CMRR is needed. This is very important in general for astronomical measurements, and in particular for measurements of the SZ signal and spectrum. 
In fact, even considering rich clusters and taking into account only the absolute emission of the CMB as a common mode signal, the CMB brightness is $\sim 10^4$ times larger than the SZ brightness from a rich cluster. This means that the CMRR of the differential instrument used to perform these measurements has to be significantly higher than $10^4$, i.e. $CMRR \gg 40 dB$ to get accurate measurements of the SZ. The requirement is even more demanding when the effects of instrument, atmospheric emission, and in general all the relevant foregrounds are included as common mode signals.

During the measurements of the SZ effect in rich clusters of galaxies, the spectrometer does not observe just the CMB: there are additional common-mode sources which we list and model below. 

As a first approximation for the astrophysical foregorunds in the mm-wave frequency range of interest here, one has to consider the brightness of diffuse interstellar dust. Its specific brightness, in the Rayleigh-Jeans region of interest here and in high Galactic latitude regions, is well approximated by:

$$I=A\left(\frac{\nu}{\nu_0}\right)^4$$

where A is normalized so that the emission at $\nu_0 = 150GHz$ is $6 \cdot 10^{-24} W/m^2  sr Hz $, which corresponds to a temperature variation of the CMB of $\sim 1\mu K$  (see  \cite{2001ApJ...553L..93M}, \cite{2015arXiv150201588P})

In the case of a room-temperature DFTS (as in OLIMPO), its thermal emission can be modeled as a grey body at room temperature ($\sim220K$ at 40 km of altitude). Due to the presence of many mirrors and optical elements (in case of the OLIMPO DFTS, 18 mirrors, 4 wire-grids and one HDPE lens) the emissivity of this grey body is not very small. In the literature we find an emissivity of $0.3\%$ \cite{bock1995emissivity} for well polished aluminum mirrors at 150 GHz; the primary mirror and secondary mirror of OLIMPO have a somewhat rougher surface, so it is conservative to assume $\sim 1\%$ emissivity; the HDPE lens has an emissivity $\sim 1\%$ and the wire-grids $\lsim 2.5\%$ \cite{2013InPhT..58...64S}.

Residual atmospheric emission is important too at these wavelengths, due to rotational transitions of water vapor, oxygen and ozone. Even working at balloon altitude, the residual emission is of the order of 0.1$\div$1\% of the 220K blackbody, so is comparable to the thermal emission of the instrument. 

\begin{figure}[hbtp]
  \centering
  \includegraphics[scale=0.35]{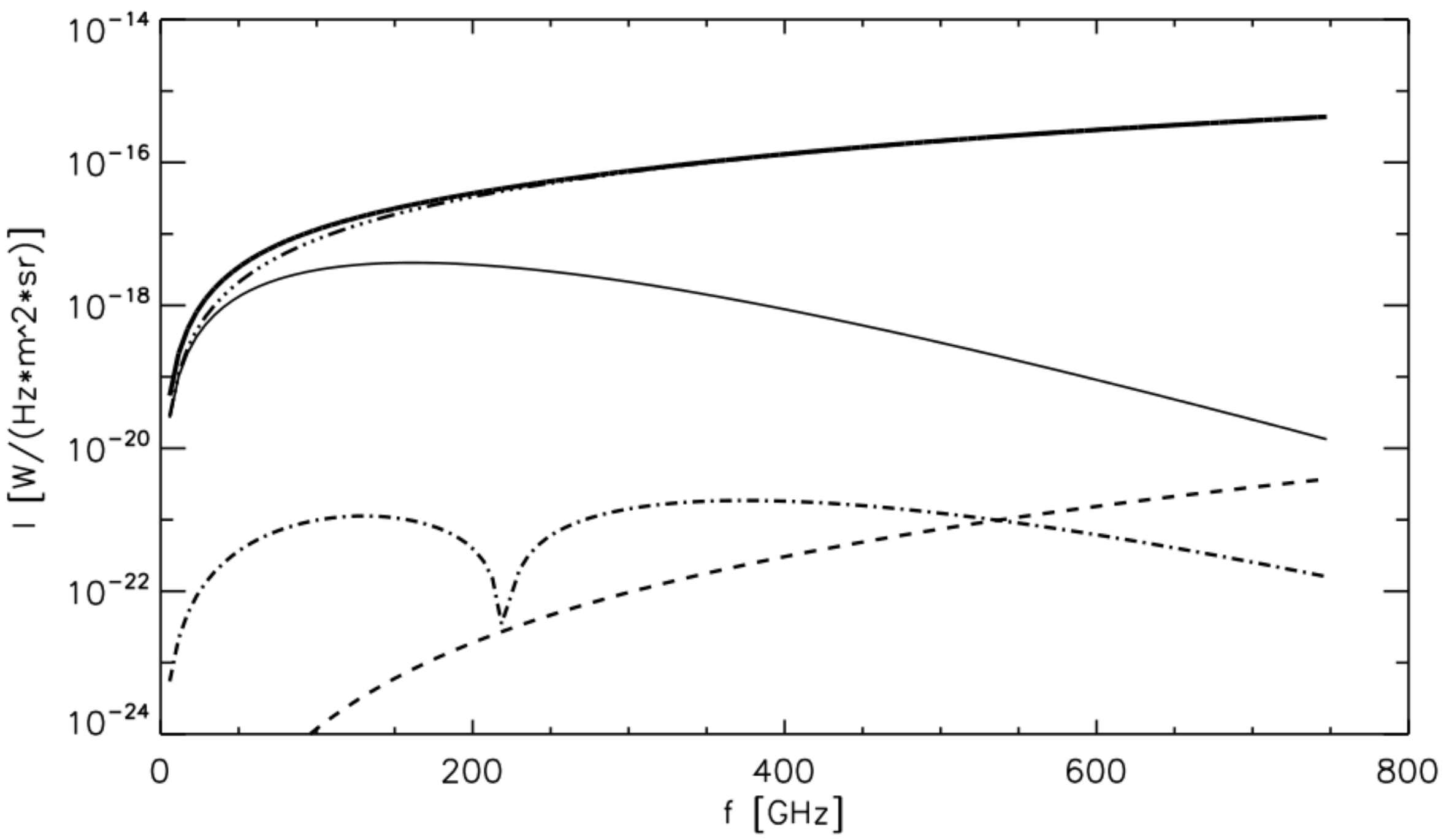}\\
  \caption{Signal and foregrounds in a typical SZ spectrum. The dot-dashed line is the SZ from a typical rich cluster; the dashed line is the brightness of interstellar dust; the thin continuos line is the absolute emission of the CMB; the three dots-dashed line is instrument emission for a complex room-temperature MPI similar to the OLIMPO one, and the thick continuos line is the total brgihtness reaching the instrument. The target signal is evidently sub-dominant with respect to the common-mode foregrounds. }\label{10.1}
\end{figure}

In \figurename \ \ref{10.1} we report a summary plot of the signals described above.

\section{Set-up description}
For this study, the OLIMPO Differential Fourier Transfor Spectrometer (DFTS)  was coupled to the OLIMPO 350 GHz detectors \cite{Coppoproc} at the output, and to custom built blackbody (BB) calibrators at the two inputs. In the following we describe these subsystems.

\subsection{The OLIMPO DFTS MPI}

\begin{figure}[hbtp]
  \centering
  \includegraphics[scale=0.3]{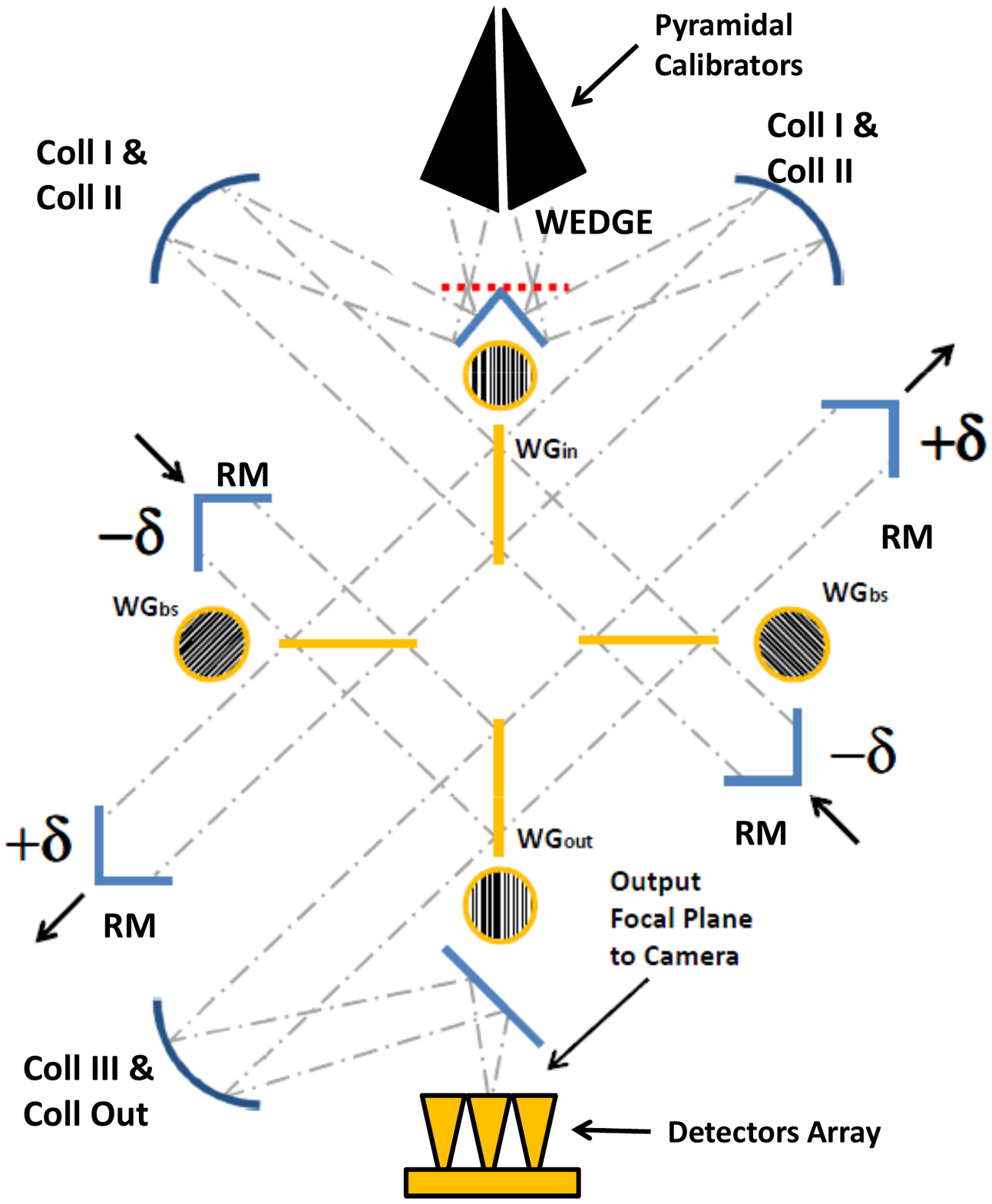}\\
  \caption{Block diagram of the double differential MPI used to measure the CMRR, including the calibration blackbody sources and the array of dectors}\label{4}
\end{figure}

A diagram of the instrument used for our measurements is shown in \figurename \ \ref{4}.

This is a double Martin-Puplett Interferometer, which was custom designed for the OLIMPO experiment, and is optimized to feed the OLIMPO cryostat and detection system \cite{OLIMPO2003}. In a classic single MPI, half of the incoming power is lost at the first wire grid. In a double MPI \cite{CARLI81} this power is recovered by a second interferometer that works synchronously with the first one, allowing for total conservation of the incoming power all the way to the detector. This feature is particularly important when observing low-brightness astrophysical sources. 

Looking at the diagram in fig. \ref{4}, the input power from two pyramidal calibrators is selected with the correct polarization by the first wire grid, to feed both the MPIs composing the instrument. After the input wire grid, half of the power travels inside the right MPI and half in the left MPI.  Here, as in a classical interferometer, we offset the retroreflectors (in this case roof mirrors) in order to introduce a delay between beams travelling along the two optical paths. In our particular design we move both retroreflectors for each MPI, which maximizes the achievable optical path difference (OPD). This is directly related to the spectral resolution of a interferometer: with 40mm of mechanical travel we get 1.8 GHz resolution. The last section of the instrument recombines the signals on the detectors. The output wire grid, with the same orientation as the input one, selects the correct polarization from each MPI to add their power. The instrument concept just described was implemented in a complex optical design using ZEMAX software, to fit the available room behind the OLIMPO primary mirror, feed the OLIMPO cryostat, and control aberrations over the entire field of view \cite{2014A&A...565A.125S}.


\subsection{Blackbody sources design}

Custom blackbody sources were developed, to be placed as close as possible to the input wedge of the DFTS, so that their emission fills homogeneously the focal plane arrays. The shape of these sources has been optimized to fill the available volume in the input section of the DFTS, while maximizing the emissivity via multiple internal reflections on high emissivity surfaces. We used a pyramidal geometry for the blackbody cavities, with a special off-axis shape to fit the available room: the pyramids are crooked and have a trapezoidal base. This shape is necessary since the input chief ray is reflected by the wedge with an angle very close to the normal, so to avoid vignetting we tilted the pyramids towards the tip of the wedge. Externally, the pyramids are made in copper, for a good heat distribution. The interior was made in $Eccosorb^{\copyright}$  CR110, which is a well documented absorber for mm-waves, often used for blackbody calibrators  \cite{Kogu04} \cite{Math99}.

The characteristics of this material are available in \cite{1985ApOpt..24.4489H}. Given the available room inside the pyramid and the absorption properties of CR110, we selected a thickness of 5mm. CR110 is available in liquid form and is hardened in the oven, so we designed and made a custom master, filling the pyramid and leaving 5mm between the master and the inner surface of the pyramid. The master was made in teflon (by Ferrotto Design snc in Pinerolo (Turin, Italy)) to easy its removal after hardening of the CR110. After baking (12 hours at $75C$), the master was removed, resulting in a pyramid made in copper externally and lined with 5mm of eccosorb CR110 internally. 

\subsection{Blackbody sources temperature control}

\begin{figure}[hbtp]
  \centering
  \includegraphics[scale=0.25]{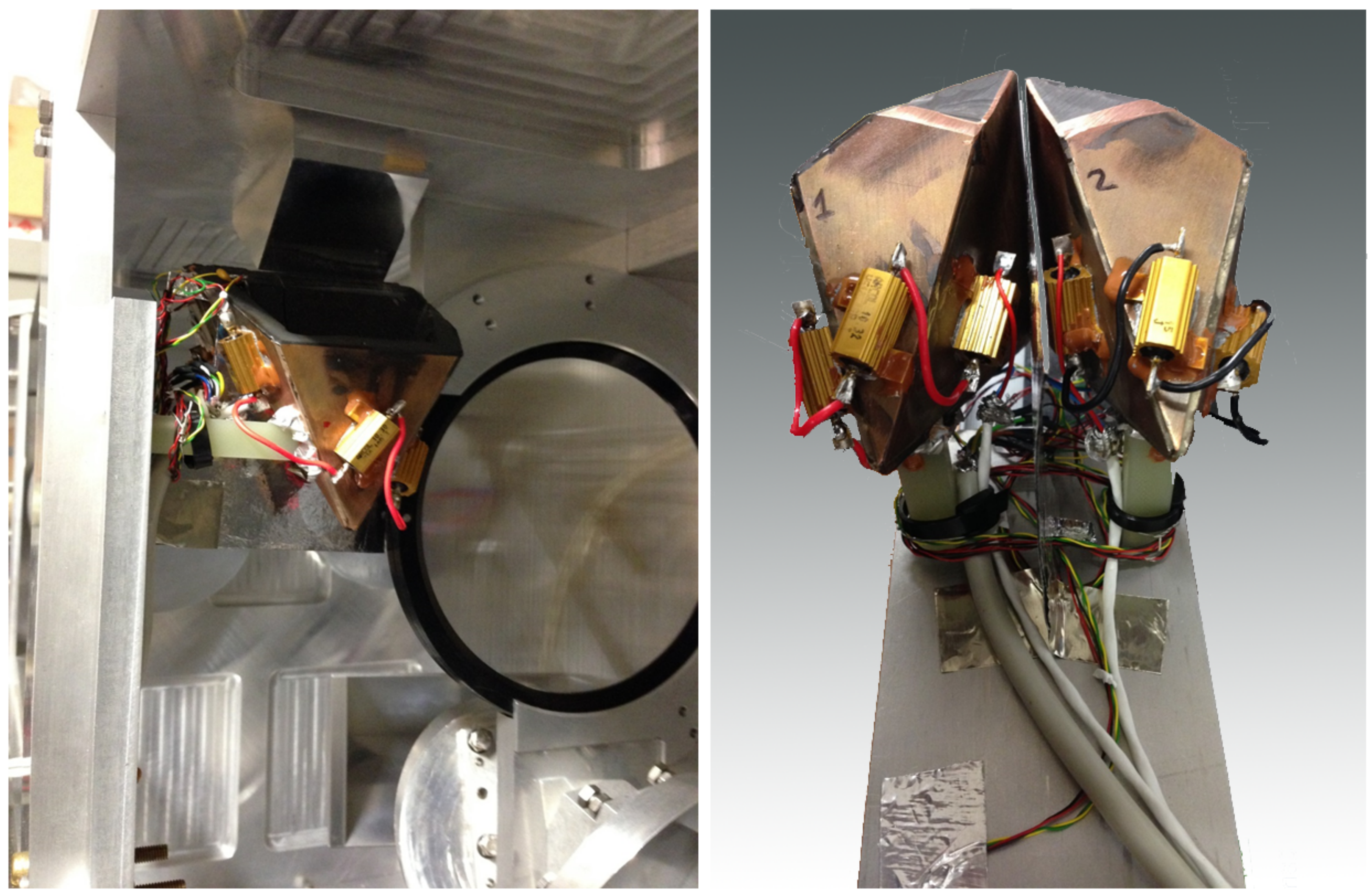}\\
  \caption{Left: Pyramidal Blackbody sources placed in front of the wedge during the measurements;
Right: A different view of the two blackbody sources, with the heaters mounted on the copper layers.}\label{10.8}
\end{figure}

On each side of the pyramidal BB we have mounted a $150\Omega$ power resistor heater, able to dissipate up to 10W. The 4 heaters, glued with a two component epoxy, are connected in series. The thermometer, a Class A Pt100 thermistor (model 362-9907  from RS components), was inserted in a small hole in the absorber and glued using CR110. The pyramids were supported by fiberglass legs, acting as insulating interfaces towards the optical bench, and shaped to place the pyramids in their correct position in front of the wedge. In \figurename \ \ref {10.8} we show the pyramids and their location in front of the wedge mirror. The thermistor is biased with a constant reference current of $1mA$, small enough to prevent Joule self-heating of the thermistor. We use the thermistor and the heater as the sensor and the actuator in a feedback system to set and stabilize the temperature of the BB at the desired value. The voltage across the thermometer is converted by a 16 bit ADC connected to a PC and controlled by a LabView program. This implements a PID algorithm, whose output is used to supply the heaters. The power dissipated in the heaters is controlled by a variable duty-cycle square wave signal (PWM) produced by a power-MOSFET. With a dynamic range of 5V, the resolution of the voltage measurement is $5000/65535=0.08mV/bit$ which corresponds, given the calibration curve of the thermometer, to a resolution in temperature $0.2K/bit$. This resolution is improved to $2mK$ by reading and averaging 10000 samples per second, so, assuming very good repeatibility of the Pt100 sensor, we are able to reproduce the same temperature difference in different runs of the measurements, within $\sim 4 mK$. The total error for the temperature measurements, instead, is dominated by the calibration uncertainty of the Pt100, and is of the order of 0.15 K.

\subsection{Thermal simulation}

\begin{figure}[hbtb]
\centering
{\includegraphics[scale=0.2]{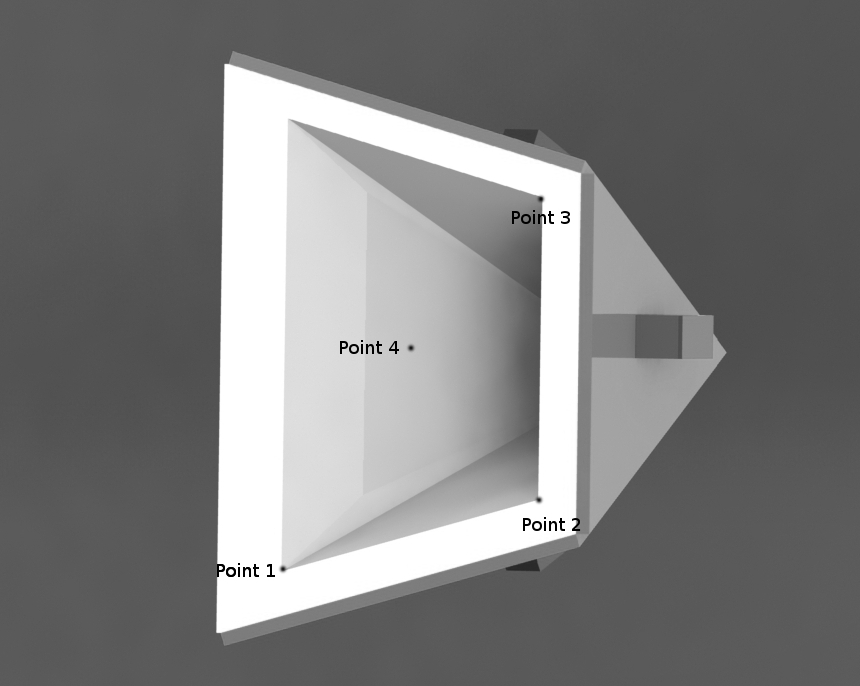}} \quad
{\includegraphics[scale=0.3]{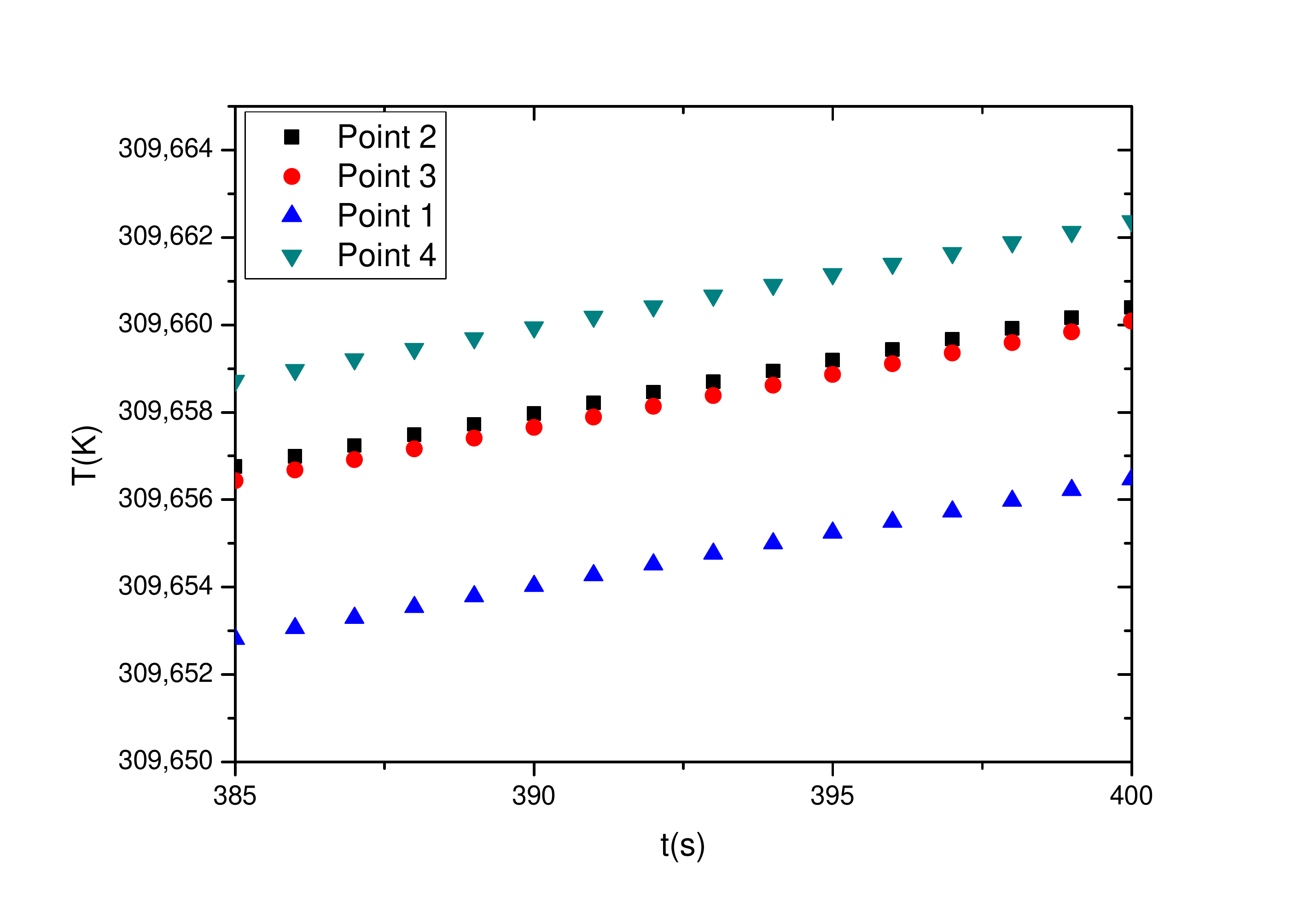}}
\caption{\emph{Top: A render of the pyramidal blackbody source, showing the location of four reference positions}; \emph{Bottom: Temperature trends at the four reference positions as obtained from a finite elements simulation. The temperature differences are $\lsim 5mK$.}} \label{10.7}
\end{figure}

An important requirement for our blackbody sources is that they are isothermal, so that the measurement with a single thermometer is representatitive of the temperature of the entire radiating surface. To this purpose the surfaces of our sources include a 2 mm thick OFHC Cu plate which ensures a very good uniformity of the temperature of the CR110 emitter coating. We have studied this problem by means of detailed finite-elements simulations.\footnote{The thermal simulation is performed using Comsol Multiphysics.}. In the simulation the pyramid surfaces are in contact with still air at ambient temperature and pressure, and transfer heat to it; in addition, the eccosorb CR100 emits like a blackbody. The thermal conductivity of eccosrb CR110 is taken from \cite{Halpern86}. The location of the heaters is also accurately replicated in the simulation. As evident from the simulation results shown in \figurename \ \ref{10.7} the temperature reached by points close to the heater and distant from it is the same within $5mK$. This level of thermal gradients across the source is negligible for our purposes. 

\subsection{Detector system}

We have used the OLIMPO receiver described in \cite{Coppoproc} and in particular the detectors for the 350 GHz band. These bolometric detectors are sensitive to a band centered at 345 GHz, with a FWHM of 30 GHz. The array overfills the corrected focal plane of the DFTS, so we used only the 13 center detectors for these measurements. To limit the radiative background and avoid saturation, a 2K neutral density filter (with 1\% transmission) was introduced in the light path to the detectors inside the cryostat. The NEP of the detectors remains anyway low enough to have a signal to noise ratio of 20 for a temperature difference of 10K.

\section{Measurements and data analysis}

Measurements were taken for different temperature differences, between -30K and +30K, as reported in Table 1.

\begin{center}
\begin{tabular}{|c|c|c|c|}
  \hline
  $T_1$ (C) & $\sigma T_1$ (C) & $T_2$ (C) & $\sigma T_2(C)$ \\
  \hline
 57.760 &   0.002 & 29.357 &   0.003\\
48.491  &  0.002 & 28.167   & 0.005\\
39.182  &  0.005 & 27.93   &  0.01\\
34.485  &  0.004  & 27.901   & 0.007\\
31.097  &   0.09  & 27.77   &  0.09\\
27.942  &  0.004  & 30.732   & 0.006\\
27.965  &  0.004  & 33.591   & 0.007\\
27.895  &  0.009 & 39.1358   &   0.009\\
28.203  &  0.006 & 47.82   &  0.01\\
28.484  &  0.007 & 56.32   &  0.01\\
  \hline
\end{tabular}

Table 1: Temperatures of the two blackbody sources during the measurements. The errors are statistical only. The total error, $\sim 0.15 K$, is dominated by the calibration error of the Pt100.
\end{center}

\begin{figure}[hbtp]
  \centering
  \includegraphics[scale=0.20]{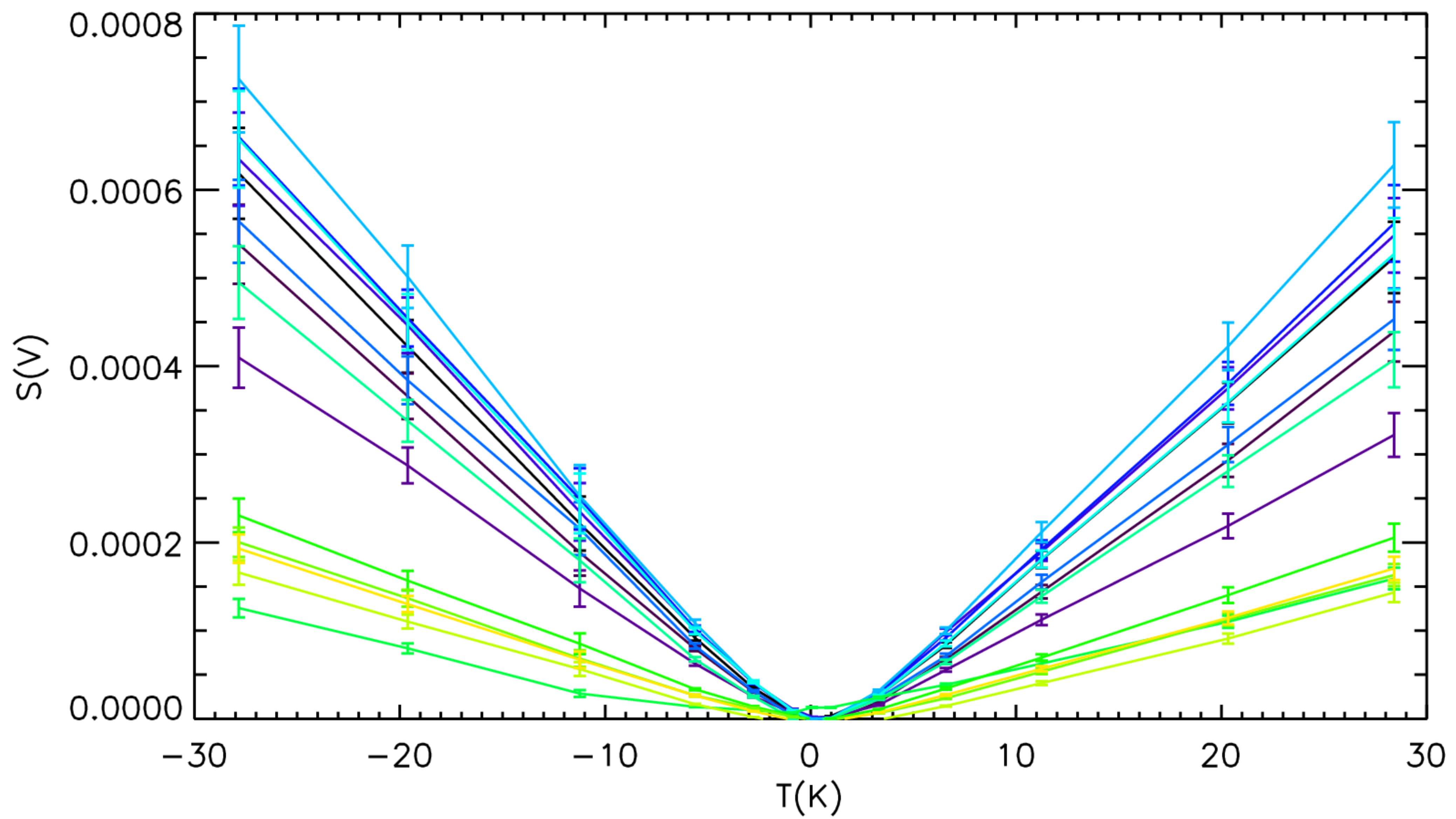}\\
  \caption{Zero-OPD interferometer peak amplitudes vs. black-body temperature difference, for all the bolometers used in this measurement.}\label{10.11}
\end{figure}

The measurements were carried out during the calibration campaign of the OLIMPO experiment, in preparation of its launch (June 2014). For each setting listed in table 1 we meaured the interferograms with all the 350 GHz detectors of the OLIMPO receiver. The measurements include a total of 4.5 hours of integration for 13 bolometers. In \figurename \ref{10.11}, for all detectors, we report the amplitudes of the zero OPD peak (which is proportional to the total power integrated over the  bandwidth of the detector), versus the difference of the temperatures of the two blackbodies. The relative error for the zero OPD signal is always much larger than the relative error in the temperature difference. 

From the plots in \figurename \ref{10.11} we derived the optical responsivity of each detector, in V/K, as the slope of best fit line. Once calibrated, the data from the different detectors were averaged together to better estimate the CMRR. 


If we model the spectrometer operation including the effect of a slight unbalance of the two inputs ports and the effect of a real common mode response, the output signal in the RJ region is
\begin{equation}\label{111}
S = R_1 T_1 - R_2 T_2 + C (R_1 T_1 + R_2 T_2)  
\end{equation} 
where $C \ll 1$ is  the inverse of the CMRR, and $R_1$ and $R_2$ are the responsivities of the detector + spectrometer system for incoming signals, at ports 1 and 2 respectively. 
If we define the unbalance factor as $\Delta R = R_2 - R_1$, with $\Delta R \ll R_1, R_2$ , we get to first order
$$
S = R_1 \bigl[ (T_1 - T_2) - {\Delta R \over R_1} T_2 + 2C T_1  \bigr]
$$
so a best fit to the data in the form

$$
S = a (T_1 - T_2) +b T_2 + c  T_1
$$

provides $R_1 = a$, ${\Delta R \over R_1} = -b/a$ and $C = c/2a$, i.e. allows us to separate the effects of real common mode and unbalance.

Note that the zero crossing signal in the plot $S$ vs $(T_1 - T_2)$ is

$$
S_0 = R_1 \bigl[ {(R_1-R_2) \over R_1} + 2C \bigr] T_{12}
$$
and depends on both ${(R_1-R_2) \over R_1}$ and $C$, so cannot be used by itself to separate the two effects.

From a linear fit to our data using equation \ref{111},  including the errors on $S$, $T_1$, $T_2$, we obtain: 
$$ R_1=(-2.41 \pm 0.26) mV/K$$
$$ R_2=(-2.79 \pm 0.12) mV/K $$
$$ C= (-5.2 \pm 0.8 ) 10^{-6} $$

Taking these results at face value, our estimate of the unbalance factor is $\Delta R =(-0.38 \pm 0.29) mV/K$. This detection of unbalance does not have a high statistical significance. If really present, could be caused by an optical misalignment of the input section of the DFTS, small orientation errors in the beamsplitters, and/or a slightly different reflectivity of two faces of the input wedge, or/and even a slight difference in the emissivity of the two blackbodies. If present, this unbalance is of the order of 15\% , so we'll carry out dedicated measurements in order to set a lower limit for this systematic effect. In \cite{2014A&A...565A.125S} we have discussed the importance of the collimation method and its critical issues, and analyzed the problems related to this misalignment.

The CMRR ($=1/|C|$) is very high, 53 dB, and is detected with good confidence. However, we cannot exclude  that better alignment of the optical components and of the beamsplitters in other instruments would result in a higher CMRR, so we prefer to consider this as a lower limit for the CMRR of MPIs. 

\begin{figure}[tb]
\centering
{\includegraphics[scale=0.3]{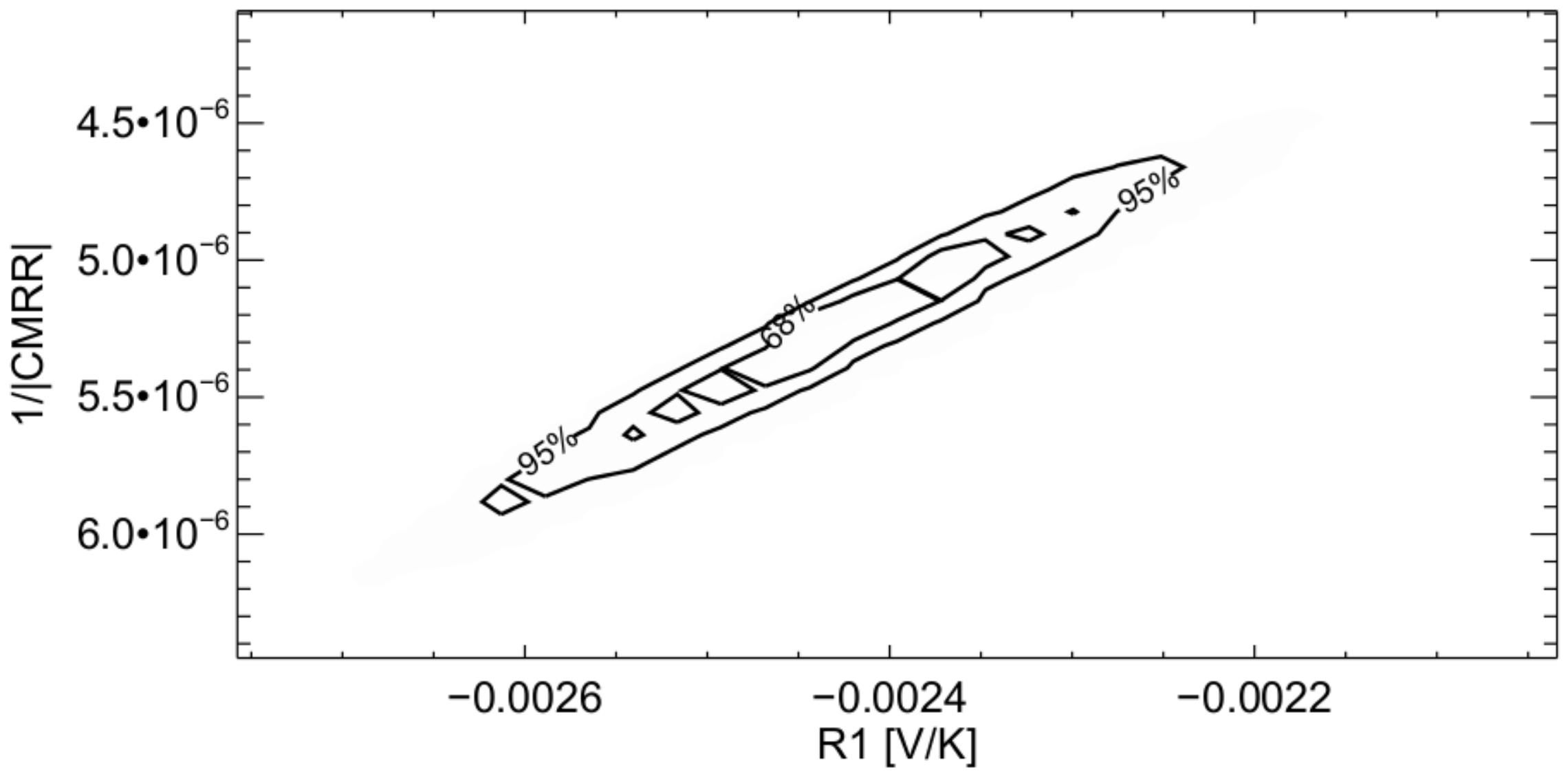}} \quad
{\includegraphics[scale=0.3]{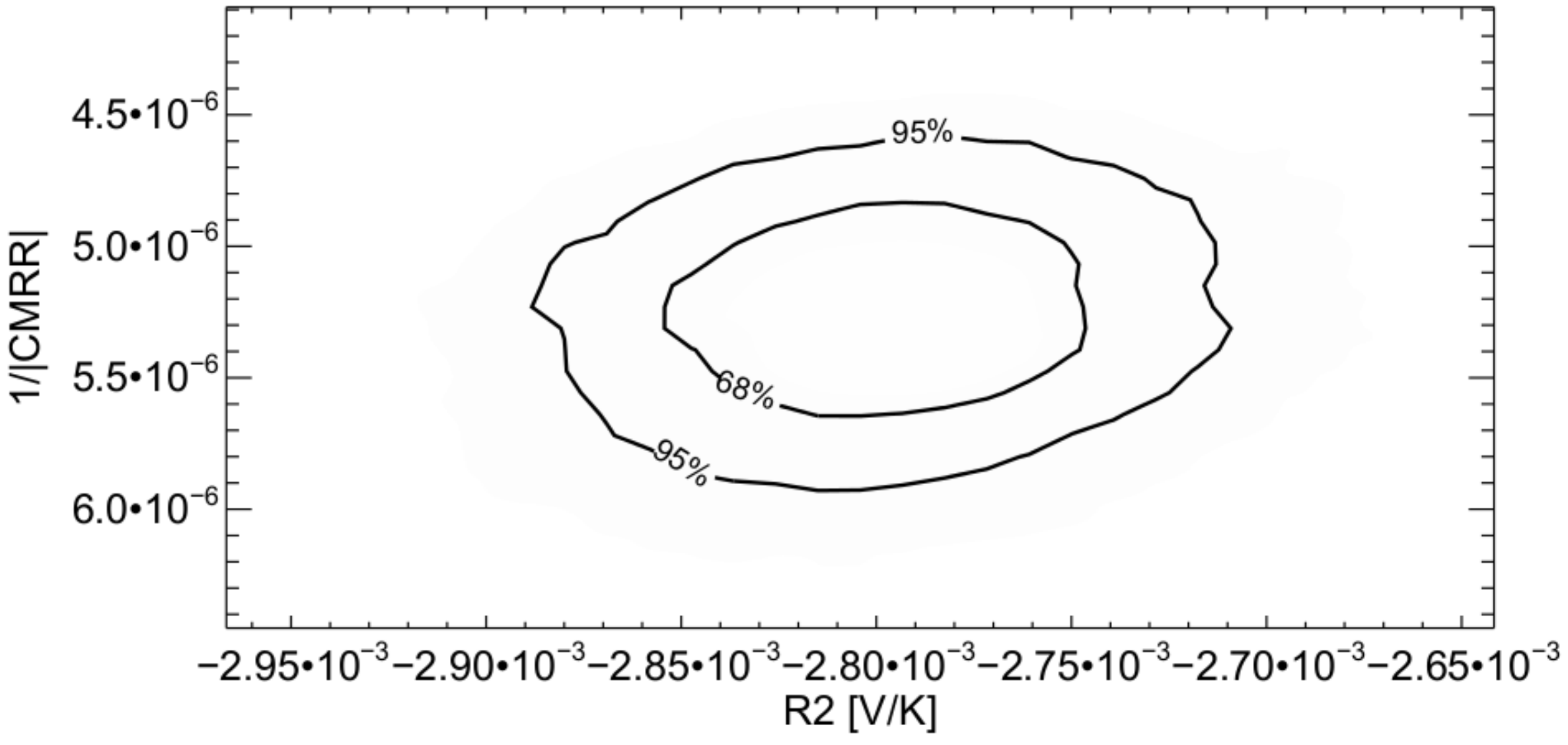}}
{\includegraphics[scale=0.3]{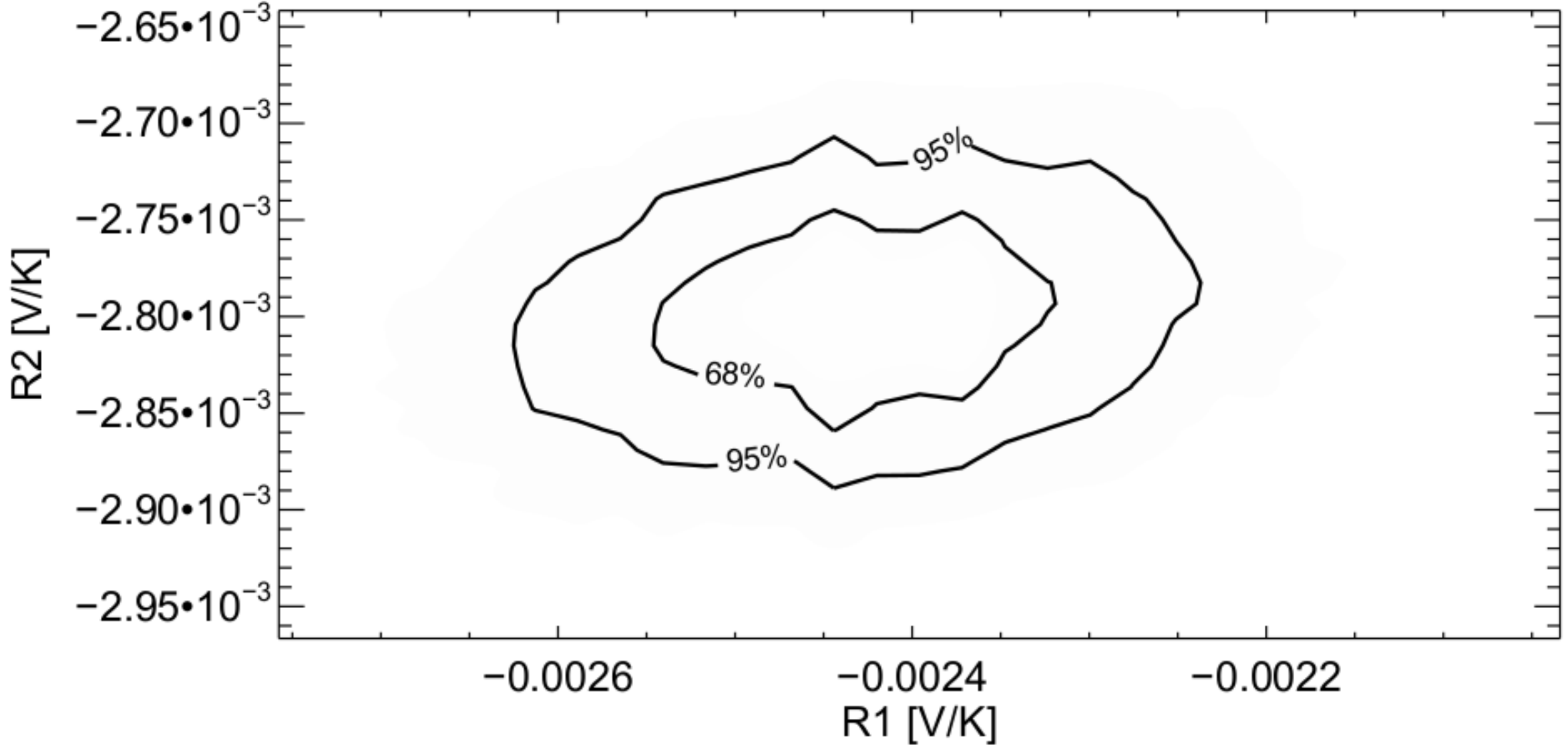}}
\caption{Contour plot of probability distribution for the three parameters: R1, R2 and the inverse of the CMRR} \label{11.0}
\end{figure}

In order to study possible correlations between the best fit parameters, we analyzed their likelihood, performing a set of Montecarlo simulations. We generated 10000 sets of measurements, with Gaussian distributions centered on the theoretical values $S_i = R_1 (T_1)_i - R_2 (T_2)_i + C (R_1 (T_1)_i + R_2 (T_2)_i)$ where  $(T_1)_i$ and $(T_2)_i$ are the temperatures listed in table 1, and $R_1, R_2$, $C$ are the best fit values listed above, with standard deviations of the Gaussian distributions equal to the measured ones. The joint probability for the different combinations of parameters are shown in  \figurename \ \ref{11.0}. There is a positive correlation between $C$ and $R_1$. This is taken into account in our estimates of the errors reported above.

\section{Discussion}

\begin{figure}[hbtp]
  \centering
  {\includegraphics[scale=0.30]{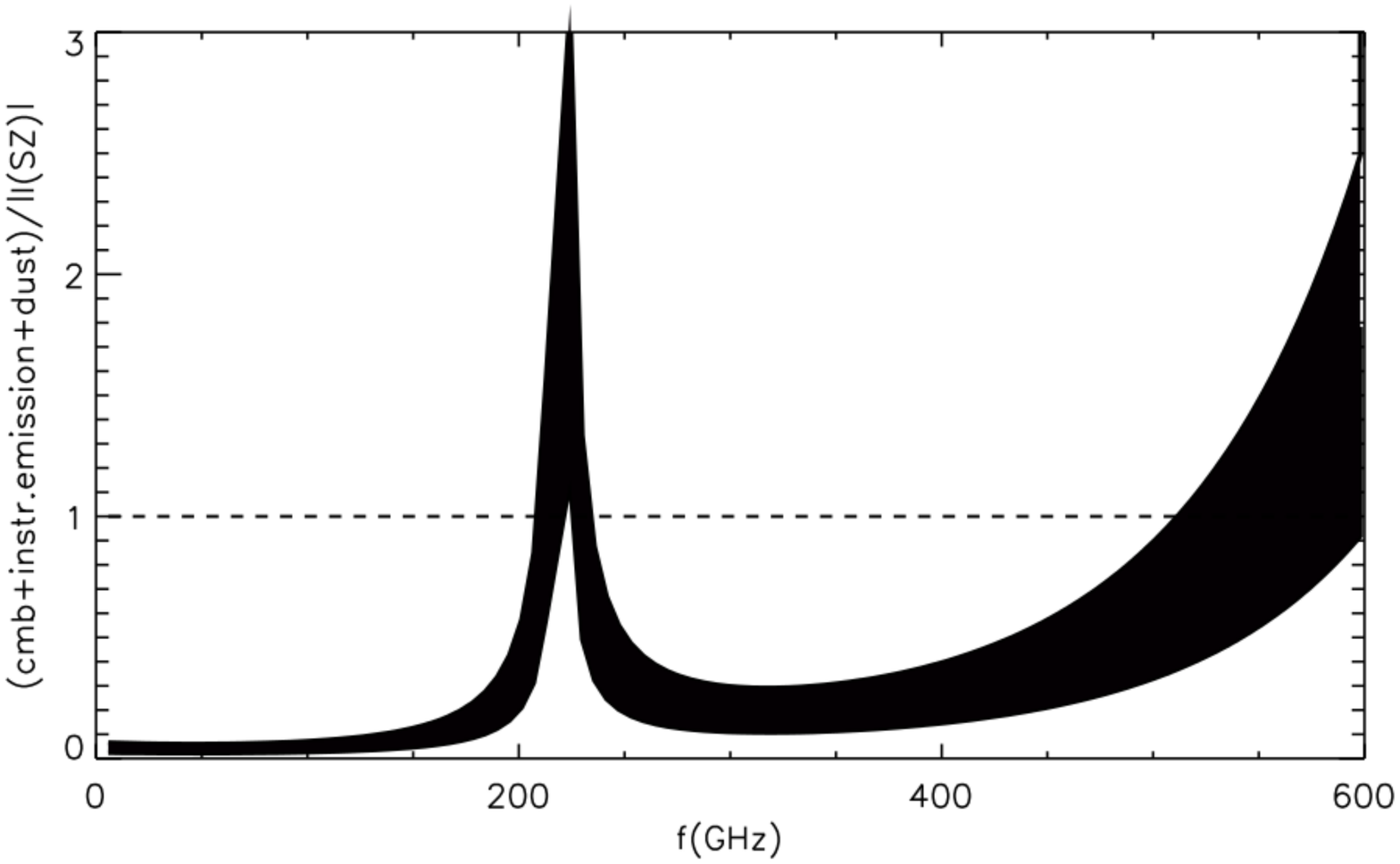}} \quad 
  {\includegraphics[scale=0.3]{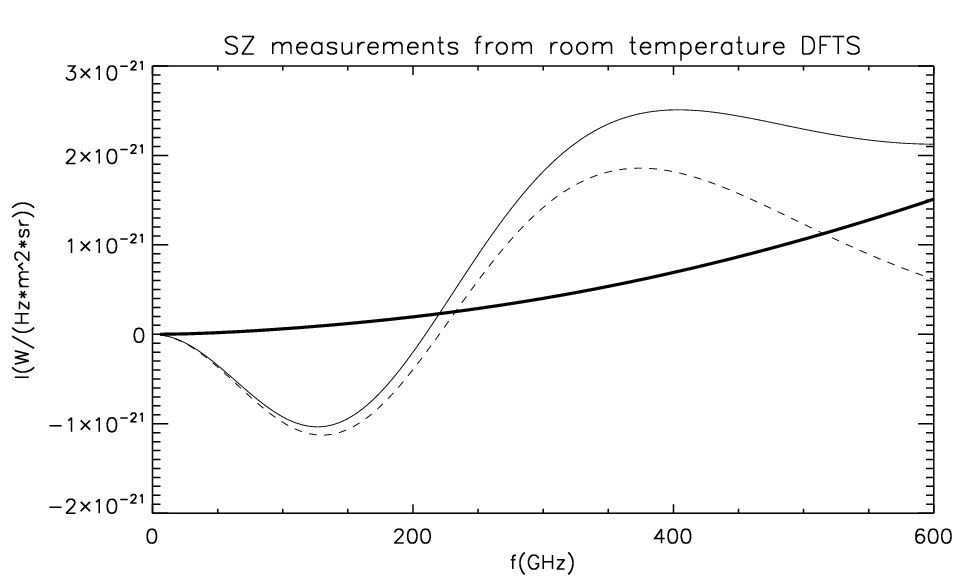}}
  \caption{\emph{Top:} Ratio between our best estimate of the common mode residual and the expected SZ signal for OLIMPO observations of a rich cluster of galaxies (black-filled band). The ratio is always well below the reference line in the OLIMPO bands; this means that it is possible to separate successfully the SZ signal from the common mode residual during the data analysis. \emph{Bottom}: Simulated measurement of the SZ spectrum obtained from the OLIMPO DFTS (at room temperature). The thickest continuos line is the common
mode residual, the dot line is the SZ signal, and the thin continuos line is the sum of the two.}\label{10.13}
\end{figure}

In order to quantify the impact of the detected CMRR on measurements of the SZ effect in rich clusters, we have used the estimates of \figurename \ref{10.1} to compute the ratio between the residual common mode signal and the expected SZ signal. For accurate measurements this ratio should be $\ll 1$. The results of this calculation are plotted in \figurename \ \ref{10.13} \emph{Top}. 

This demonstrates that, despite of the fact that the DFTS operates at room temperature, OLIMPO can detect the SZ spectrum of a typical rich cluster with a common mode contamination smaller than the signal to be measured. We can simulate a SZ measurement using such result, assuming that the common mode is proportional to the common mode brightness, with an average amplitude integrated over the bandwidth of $(1.5 \pm 0.1)mK$. Assuming that the common mode residual is simply proportional to the input power, we show in  \figurename \ \ref{10.13} \emph{bottom} the result of this simulation, for the OLIMPO room-temperature DFTS. 

\begin{figure}[tb]
\centering
\includegraphics[scale=0.3]{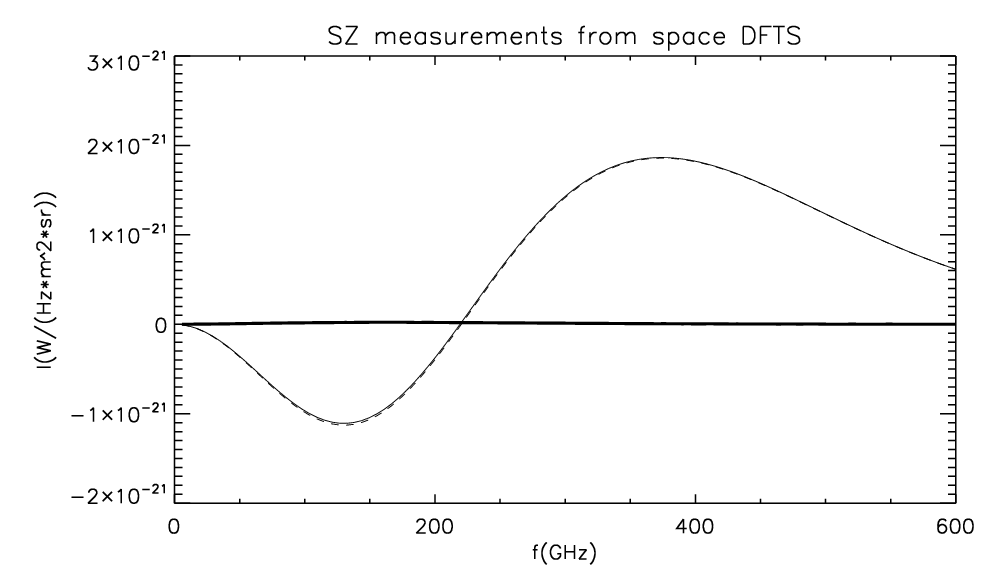}
\caption{Simulated measurement of the SZ spectrum obtained from a cryogenic space-borne DFTS. Lines as in \figurename \ \ref{10.13} \emph{bottom}.  The effects of common mode are totally negligible due to the low radiative background environment.} \label{10.14}
\end{figure}

This exercise has been repeated for a cryogenic implementation of the same DFTS, to be used, for example, in the Millimetron \cite{milli} mission. Given the low radiative background in the cryogenic environment, the common mode residual becomes negligible (of the order of $\sim 1\%$ when averaged over the bandwidth) with respect to the spectrum of the SZ effect, as shown in figure \figurename \ \ref{10.14}.

\section{Conclusions}

We used the OLIMPO DFTS as an example of large-throughput, room temperature MPI, optimized for measurements of the Sunyaev-Zeldovich effect, a demanding application where the CMRR has to be very high. We built two custom black body calibrators to measure the CMRR. The two calibrators were designed to fit the two input ports of the DFTS, and their temperature was controlled with $2 mK$ stability and measured with 0.15 K accuracy. During the flight preparation campaign for the OLIMPO launch we carried out spectral measuements for different temperatures of the blackbody sources, and evaluated both the {\sl real} common mode residual and the unbalance for the DFTS. We set a lower limit for the {\sl real} CMRR of $\gsim 50dB$, separating this effect from input unbalance. We conclude that, if properly balanced, this instrument is able to measure the SZ effect in rich clusters of galaxies with a reasonable offset, which can be removed in data analysis. Moreover, we have shown that a cryogenic implementation of this instrument is suitable for space-based measurements of the faintest SZ clusters, as planned in the Millimetron mission.

\section{Acknowledgments}
This work has been supported by ASI (Agenzia Spaziale Italiana) grants OLIMPO and Millimetron, and by PNRA (Italian National Antarctic Research Program) grant BRAIN.
We warmly acknowledge Mr. Giorgio Amico for carefully machining many parts of the experiment setup.

\bibliography{sample}

\begin{thebibliography}{10}
\newcommand{\enquote}[1]{``#1''}

\bibitem{Mart70}
D.~H. {Martin} and E.~{Puplett}, \enquote{{Polarised interferometric
  spectrometry for the millimeter and submillimeter spectrum.}} Infrared
  Physics \textbf{10}, 105--109 (1970).

\bibitem{Math93}
J.~C. {Mather}, D.~J. {Fixsen}, and R.~A. {Shafer}, \enquote{{Design for the
  COBE far-infrared absolute spectrophotometer (FIRAS)},} in \enquote{Infrared
  Spaceborne Remote Sensing,} , vol. 2019 of \emph{Society of Photo-Optical
  Instrumentation Engineers (SPIE) Conference Series}, M.~S. {Scholl}, ed.
  (1993), vol. 2019 of \emph{Society of Photo-Optical Instrumentation Engineers
  (SPIE) Conference Series}, pp. 168--179.

\bibitem{Grif07}
M.~{Griffin}, A.~{Abergel}, P.~{Ade}, P.~{Andr{\'e}}, J.-P. {Baluteau},
  J.~{Bock}, A.~{Franceschini}, W.~{Gear}, J.~{Glenn}, D.~{Griffin}, K.~{King},
  E.~{Lellouch}, S.~{Madden}, D.~{Naylor}, S.~{Oliver}, G.~{Olofsson},
  M.~{Page}, I.~{Perez-Fournon}, M.~{Rowan-Robinson}, P.~{Saraceno},
  E.~{Sawyer}, B.~{Swinyard}, L.~{Vigroux}, G.~{Wright}, and {SPIRE
  Consortium}, \enquote{{The Herschel-SPIRE instrument and its capabilities for
  extragalactic astronomy},} Advances in Space Research \textbf{40}, 612--619
  (2007).

\bibitem{Math90}
J.~C. {Mather}, E.~S. {Cheng}, R.~E. {Eplee}, Jr., R.~B. {Isaacman}, S.~S.
  {Meyer}, R.~A. {Shafer}, R.~{Weiss}, E.~L. {Wright}, C.~L. {Bennett}, N.~W.
  {Boggess}, E.~{Dwek}, S.~{Gulkis}, M.~G. {Hauser}, M.~{Janssen},
  T.~{Kelsall}, P.~M. {Lubin}, S.~H. {Moseley}, Jr., T.~L. {Murdock}, R.~F.
  {Silverberg}, G.~F. {Smoot}, and D.~T. {Wilkinson}, \enquote{{A preliminary
  measurement of the cosmic microwave background spectrum by the Cosmic
  Background Explorer (COBE) satellite},} Ap.j \textbf{354}, L37--L40 (1990).

\bibitem{Grif10}
M.~J. {Griffin}, A.~{Abergel}, A.~{Abreu}, P.~A.~R. {Ade}, P.~{Andr{\'e}},
  J.-L. {Augueres}, T.~{Babbedge}, Y.~{Bae}, T.~{Baillie}, J.-P. {Baluteau},
  M.~J. {Barlow}, G.~{Bendo}, D.~{Benielli}, J.~J. {Bock}, P.~{Bonhomme},
  D.~{Brisbin}, C.~{Brockley-Blatt}, M.~{Caldwell}, C.~{Cara},
  N.~{Castro-Rodriguez}, R.~{Cerulli}, P.~{Chanial}, S.~{Chen}, E.~{Clark},
  D.~L. {Clements}, L.~{Clerc}, J.~{Coker}, D.~{Communal}, L.~{Conversi},
  P.~{Cox}, D.~{Crumb}, C.~{Cunningham}, F.~{Daly}, G.~R. {Davis}, P.~{de
  Antoni}, J.~{Delderfield}, N.~{Devin}, A.~{di Giorgio}, I.~{Didschuns},
  K.~{Dohlen}, M.~{Donati}, A.~{Dowell}, C.~D. {Dowell}, L.~{Duband},
  L.~{Dumaye}, R.~J. {Emery}, M.~{Ferlet}, D.~{Ferrand}, J.~{Fontignie},
  M.~{Fox}, A.~{Franceschini}, M.~{Frerking}, T.~{Fulton}, J.~{Garcia},
  R.~{Gastaud}, W.~K. {Gear}, J.~{Glenn}, A.~{Goizel}, D.~K. {Griffin},
  T.~{Grundy}, S.~{Guest}, L.~{Guillemet}, P.~C. {Hargrave}, M.~{Harwit},
  P.~{Hastings}, E.~{Hatziminaoglou}, M.~{Herman}, B.~{Hinde}, V.~{Hristov},
  M.~{Huang}, P.~{Imhof}, K.~J. {Isaak}, U.~{Israelsson}, R.~J. {Ivison},
  D.~{Jennings}, B.~{Kiernan}, K.~J. {King}, A.~E. {Lange}, W.~{Latter},
  G.~{Laurent}, P.~{Laurent}, S.~J. {Leeks}, E.~{Lellouch}, L.~{Levenson},
  B.~{Li}, J.~{Li}, J.~{Lilienthal}, T.~{Lim}, S.~J. {Liu}, N.~{Lu},
  S.~{Madden}, G.~{Mainetti}, P.~{Marliani}, D.~{McKay}, K.~{Mercier},
  S.~{Molinari}, H.~{Morris}, H.~{Moseley}, J.~{Mulder}, M.~{Mur}, D.~A.
  {Naylor}, H.~{Nguyen}, B.~{O'Halloran}, S.~{Oliver}, G.~{Olofsson}, H.-G.
  {Olofsson}, R.~{Orfei}, M.~J. {Page}, I.~{Pain}, P.~{Panuzzo},
  A.~{Papageorgiou}, G.~{Parks}, P.~{Parr-Burman}, A.~{Pearce}, C.~{Pearson},
  I.~{P{\'e}rez-Fournon}, F.~{Pinsard}, G.~{Pisano}, J.~{Podosek}, M.~{Pohlen},
  E.~T. {Polehampton}, D.~{Pouliquen}, D.~{Rigopoulou}, D.~{Rizzo}, I.~G.
  {Roseboom}, H.~{Roussel}, M.~{Rowan-Robinson}, B.~{Rownd}, P.~{Saraceno},
  M.~{Sauvage}, R.~{Savage}, G.~{Savini}, E.~{Sawyer}, C.~{Scharmberg},
  D.~{Schmitt}, N.~{Schneider}, B.~{Schulz}, A.~{Schwartz}, R.~{Shafer}, D.~L.
  {Shupe}, B.~{Sibthorpe}, S.~{Sidher}, A.~{Smith}, A.~J. {Smith}, D.~{Smith},
  L.~{Spencer}, B.~{Stobie}, R.~{Sudiwala}, K.~{Sukhatme}, C.~{Surace}, J.~A.
  {Stevens}, B.~M. {Swinyard}, M.~{Trichas}, T.~{Tourette}, H.~{Triou},
  S.~{Tseng}, C.~{Tucker}, A.~{Turner}, M.~{Vaccari}, I.~{Valtchanov},
  L.~{Vigroux}, E.~{Virique}, G.~{Voellmer}, H.~{Walker}, R.~{Ward},
  T.~{Waskett}, M.~{Weilert}, R.~{Wesson}, G.~J. {White}, N.~{Whitehouse},
  C.~D. {Wilson}, B.~{Winter}, A.~L. {Woodcraft}, G.~S. {Wright}, C.~K. {Xu},
  A.~{Zavagno}, M.~{Zemcov}, L.~{Zhang}, and E.~{Zonca}, \enquote{{The
  Herschel-SPIRE instrument and its in-flight performance},} A$\&$A
  \textbf{518}, L3 (2010).

\bibitem{Kogu11}
C.~{A. Kogut}, {Fixsen D.T.}, \enquote{The primordial inflation
  explorer(pixie): A nulling polarimeter for cosmic microwave background
  observations,}  .

\bibitem{SZ69}
S.~{Zeldovich Y. B.}, \enquote{The interaction of matter and radiation in a
  hot-model universe,} Ap\&SS \textbf{4}, 301 (1969).

\bibitem{Chlu12}
S.~{Chluba J.}, { Fung J.}, \enquote{Radiative transfer effects during
  primordial helium recombination,} MNRAS \textbf{423}, 3227--3242 (2012).

\bibitem{debe12}
P.~{de Bernardis}, B.~A., and A.~e.~a. {Bardi}, \enquote{Sagace: the
  spectroscopic active galaxies and clusters explorer,} proc. of the 12th
  Marcel Grossmann Meeting on General Relativity p. 2133 (2012).

\bibitem{debe12a}
P.~{de Bernardis}, S.~{Colafrancesco}, G.~{D'Alessandro}, L.~{Lamagna},
  P.~{Marchegiani}, S.~{Masi}, and A.~{Schillaci}, \enquote{{Low-resolution
  spectroscopy of the Sunyaev-Zel'dovich effect and estimates of cluster
  parameters},} A$\&$A \textbf{538}, A86 (2012).

\bibitem{Benn94}
{Bennett C.L.}, \enquote{Morphology of the interstellar cooling lines detected
  by cobe,} ApJ .

\bibitem{Vier13}
M.~{ M. P. Viero}, { L. Wang}, \enquote{Hermes: Cosmic infrared background
  anisotropies and the clustering of dusty star-forming galaxies,} ApJ p. 772
  (2013).

\bibitem{Yu01}
J.~{ Qingjuan Yu}, { David N. Spergel}, \enquote{Rayleigh scattering and
  microwave background fluctuations,} ApJ \textbf{558}.

\bibitem{Lewi13}
{ Lewis A.}, \enquote{Rayleigh scattering: blue sky thinking for future cmb
  observations,} JCAP  (2013).

\bibitem{Shao11}
P.~{Jiawei Shao}, \enquote{The kinetic sz tomography with spectroscopic
  redshift surveys,} Mon.Not.Roy.Astron.Soc. \textbf{413}, 628--642 (2011).

\bibitem{Hina08}
J.~{Hanae Inami}, {Matt Bradford}, \enquote{A broadband millimeter-wave
  spectrometer z-spec: sensitivity and ulirgs,} SPIE Proceedings: Millimeter
  and Submillimeter Detectors and Instrumentation for Astronomy IV
  \textbf{7020} (2008).

\bibitem{Brya15}
C.~P.~M. {Sean Bryan}, { George Che}, \enquote{A compact filter-bank waveguide
  spectrometer for millimeter wavelengths,} astro-ph/1502.02735v2 .

\bibitem{DAless2015}
{G. D'Alessandro}, {P. de Bernardis}, {S.Masi}, and {A.Schillaci},
  \enquote{Effect on beamsplitter orientation error on mpi interferometer,} in
  preparation  (2015).

\bibitem{2014A&A...565A.125S}
A.~{Schillaci}, G.~{D'Alessandro}, P.~{de Bernardis}, S.~{Masi}, C.~{Paiva
  Novaes}, M.~{Gervasi}, and M.~{Zannoni}, \enquote{{Efficient differential
  Fourier-transform spectrometer for precision Sunyaev-Zel'dovich effect
  measurements},} A$\&$A \textbf{565}, A125 (2014).

\bibitem{2001ApJ...553L..93M}
S.~{Masi}, P.~A.~R. {Ade}, J.~J. {Bock}, A.~{Boscaleri}, B.~P. {Crill}, P.~{de
  Bernardis}, M.~{Giacometti}, E.~{Hivon}, V.~V. {Hristov}, A.~E. {Lange},
  P.~D. {Mauskopf}, T.~{Montroy}, C.~B. {Netterfield}, E.~{Pascale},
  F.~{Piacentini}, S.~{Prunet}, and J.~{Ruhl}, \enquote{{High-Latitude Galactic
  Dust Emission in the BOOMERANG Maps},} apjl \textbf{553}, L93--L96 (2001).

\bibitem{2015arXiv150201588P}
{Planck Collaboration}, R.~{Adam}, P.~A.~R. {Ade}, N.~{Aghanim}, M.~I.~R.
  {Alves}, M.~{Arnaud}, M.~{Ashdown}, J.~{Aumont}, C.~{Baccigalupi}, A.~J.
  {Banday}, and et~al., \enquote{{Planck 2015 results. X. Diffuse component
  separation: Foreground maps},} ArXiv e-prints  (2015).

\bibitem{bock1995emissivity}
F.~L. {Bock JJ.}, {Parikh MK.}, \enquote{Emissivity measurements of reflective
  surfaces at near-millimeter wavelengths,} Applied optics \textbf{34},
  4812--4816.

\bibitem{2013InPhT..58...64S}
A.~{Schillaci}, E.~{Battistelli}, G.~{D' Alessandro}, P.~{de Bernardis}, and
  S.~{Masi}, \enquote{{On the emissivity of wire-grid polarizers for
  astronomical observations at mm-wavelengths},} Infrared Physics and
  Technology \textbf{58}, 64--68 (2013).

\bibitem{Coppoproc}
A.~{Coppolecchia}, \enquote{Olimpo: A 4-bands detectors array for balloon-borne
  observations of the sunyaev-zeldovich effect. in: New horizons for
  observational cosmology,} Proceeding Varenna Cosmology confereces
  \textbf{186} (2013).

\bibitem{OLIMPO2003}
S.~{Masi}, P.~{Ade}, P.~{de Bernardis}, A.~{Boscaleri}, M.~{De Petris}, G.~{De
  Troia}, M.~{Fabrini}, M.~{Giacometti}, A.~{Iacoangeli}, L.~{Lamagna},
  A.~{Lange}, P.~{Lubin}, P.~{Mauskopf}, A.~{Melchiorri}, F.~{Melchiorri},
  F.~{Nati}, L.~{Nati}, A.~{Orlando}, E.~{Pascale}, F.~{Piacentini},
  M.~{Pierre}, G.~{Polenta}, Y.~{Rephaeli}, G.~{Romeo}, and D.~{Yvon},
  \enquote{{OLIMPO: A few arcmin resolution survey of the sky at mm and sub-mm
  wavelengths},} Mem.Soc.Astron.It \textbf{74}, 96 (2003).

\bibitem{CARLI81}
M.~{Carli B.}, \enquote{{Signal doubling in the Martin-Puplett
  interferometer},} International Journal of Infrared and Millimeter Waves
  \textbf{2}, 1045--1051 (1981).

\bibitem{Kogu04}
W.~F. L. M. L. S.~L. {Kogut}, {Alan}, \enquote{Design and calibration of a
  cryogenic blackbody calibrator at centimeter wavelengths,} Rev. Sci. Instrum.
  \textbf{75}, 5079--5083 (2004).

\bibitem{Math99}
S.~M.~W. {Mather}, {Fixsen}, \enquote{Calibrator design for the cobe far
  infrared absolute spectrophotometer (firas),} Astrophys. J. \textbf{512},
  511--520 (1999).

\bibitem{1985ApOpt..24.4489H}
E.~{Hemmati H.}, { Mather J.C.}, \enquote{{Submillimeter and millimeter wave
  characterization of absorbing materials},} Applied Optics \textbf{24},
  4489--4492 (1985).

\bibitem{Halpern86}
E.~V. {Mark Halpern}, {Herbert P. Gush}, \enquote{Far infrared transmission of
  dielectrics at cryogenic and room temperatures: glass, fluorogold, eccosorb,
  stycast, and various plastics,} Appl. Opt. \textbf{25}, 565--570 (1986).

\bibitem{milli}
P.~{Smirnov}, {Baryshev}, \enquote{Space mission millimetron for terahertz
  astronomy,} Society of Photo-Optical Instrumentation Engineers (SPIE)
  Conference Series \textbf{8442} (2012).

\end{thebibliography}


\end{document}